\documentclass[runningheads]{llncs}
\usepackage[utf8]{inputenc}

\usepackage{hyperref}
\usepackage{longtable}
\usepackage{subcaption}

\usepackage{listings}
\usepackage[dvipsnames]{xcolor}
\usepackage{textcomp}
\usepackage{lstcoq}

\lstdefinelanguage{michelson}{%
   columns=fullflexible,%
   basicstyle=\small\tt ,
   commentstyle=\slshape,%
   keywords={%
     \{,\},
     DROP, DUP, SWAP, PUSH, SOME, NONE, UNIT, IF_NONE,%
   PAIR, CAR, CDR, LEFT, RIGHT, IF_LEFT, IF_RIGHT, NIL,%
   CONS, IF_CONS, SIZE, EMPTY_SET, EMPTY_MAP, MAP, ITER,%
   MEM, GET, UPDATE, IF, LOOP, LOOP_LEFT, LAMBDA, EXEC,%
   DIP, FAILWITH, CAST, RENAME, CONCAT, SLICE, PACK,%
   UNPACK, ADD, SUB, MUL, EDIV, ABS, NEG, LSL, LSR,%
   OR, AND, XOR, NOT, COMPARE, EQ, NEQ, LT, GT, LE,%
   GE, SELF, CONTRACT, TRANSFER_TOKENS, SET_DELEGATE,%
   CREATE_ACCOUNT, CREATE_CONTRACT, CREATE_CONTRACT,%
   IMPLICIT_ACCOUNT, NOW, AMOUNT, BALANCE, CHECK_SIGNATURE,%
   BLAKE, SHA, SHA, HASH_KEY, STEPS_TO_QUOTA, SOURCE,%
   SENDER, ADDRESS,%
   CMPEQ,CMPNEQ,CMPLT,CMPGT,CMPLE,CMPGE,%
   IFEQ,IFNEQ,IFLT,IFGT,IFLE,IFGE,%
   IFCMPEQ,IFCMPNEQ,IFCMPLT,IFCMPGT,IFCMPLE,IFCMPGE,%
   FAIL,%
   ASSERT,%
   ASSERT_EQ,ASSERT_NEQ,ASSERT_LT,ASSERT_LE,ASSERT_GT,ASSERT_GE,%
   ASSERT_CMPEQ,ASSERT_CMPNEQ,ASSERT_CMPLT,ASSERT_CMPLE,ASSERT_CMPGT,ASSERT_CMPGE,%
   ASSERT_NONE,ASSERT_SOME,%
   ASSERT_LEFT,ASSERT_RIGHT,%
   UNPAIR,%
   },%
   alsoletter={'},
   upquote=true,
   keywordstyle={\bfseries\sffamily},%
   morekeywords=[2]{%
     key, unit, signature, option, list, set, operation, address,%
     contract, pair, or, lambda, big_map, map,%
     int, nat, string, bytes, mutez, bool, key_hash, %
     timestamp, 'a, 'b, 'S, 'p%
   },%
   keywordstyle=[2]{\bfseries\ttfamily},%
   classoffset=2,%
   morekeywords=[3]{%
     storage, parameter, code %
   },%
   keywordstyle=[3]{\bfseries},%
   sensitive,%
   comment=[l]\#,%
   morestring=[d]",%"
   literate={->}{{$\rightarrow{}$}}1%
}[keywords,comments,strings]%

\lstMakeShortInline[language=michelson,basicstyle=\ttfamily,keywordstyle=\bfseries\sffamily\small]!

\begin{document}
\title{Mi-Cho-Coq, a framework for certifying Tezos Smart Contracts}
\author{Bruno Bernardo \and
Rapha\"el Cauderlier \and
Zhenlei Hu \and
Basile Pesin \and
Julien Tesson}
\institute{Nomadic Labs, Paris, France\\ \email{\{first\_name.last\_name\}@nomadic-labs.com}}
\maketitle              % typeset the header of the contribution
%
% similar to mi_cho_coq_abstract.txt
\begin{abstract}
  Tezos is a blockchain launched in June 2018. It is written in OCaml
  and supports smart contracts. Its smart contract language is called
  Michelson and it has been designed with formal verification in
  mind. In this article, we present Mi-Cho-Coq, a Coq framework for
  verifying the functional correctness of Michelson smart contracts.
  As a case study, we detail the certification of a Multisig contract
  with the Mi-Cho-Coq framework.

\end{abstract}

\section{Introduction to Tezos}
Tezos is a public blockchain launched in June 2018. It is mostly
implemented in OCaml~\cite{leroy:ocaml4.08} and its code is open
source~\cite{tezosGitLab}. Like Ethereum, Tezos is an account based
smart contract platform. This section is a high-level
broad overview of Tezos to distinguish it from similar
projects like Bitcoin and Ethereum.

\subsubsection*{Consensus algorithm}
Unlike Bitcoin and Ethereum, Tezos' \textbf{consensus algorithm} is
based on a \textbf{Proof-of-Stake} algorithm~\cite{tezosLiquidPos}:
rights to produce new blocks are given to accounts that own a stake.
More precisely, there is a delegation mechanism and the
block-producing rights of each account are given in probabilistic proportion to the
number of tokens that have been \emph{delegated} to this account.
Block producers have to make a security deposit that is slashed if
their behaviour looks malicious, for example if they produce two
different blocks for the same level (double spending attack).
% As this subject is out of scope of this paper's topic, the interested reader is kindly invited to this reading here
% For each block level, a list of production rights is
% computed; a block of a some level is valid if it is well formed, contains only valid
% transactions and is signed by an account in the list for that level.
% The position of the account in the list
% deterministically and independently by every node. A commitment
% scheme (ref?) is used so that the computation of these rights can be
% done independently by every node
\subsubsection*{On-chain voting}
Another key point of Tezos is its \textbf{on-chain governance
  mechanism}. The codebase can be changed by a vote of the token
holders via their delegates. This helps preventing divisions amongst
the community that could lead to forks. Only a delimited part of the
codebase, named the \emph{economic ruleset} or the \emph{economic
  protocol}~\cite{goodman2014positionpaper,goodman2014whitepaper}, can
be changed. This part contains the rules that define what a valid
transaction is, what a valid block is, as well as how to choose
between multiple chains. %
Thus, the economic ruleset contains, amongst other things, the
consensus algorithm, the language for smart contracts and also
the voting rules~\cite{tezosVoting}. It does
not contain the network and storage layers.
If a proposal is accepted, nodes need not to stop and restart:
the new code is downloaded from other peers, dynamically compiled and hot swapped.
At the moment, the voting procedure lasts approximately three months but that could
be changed in the future via a vote.

% NOTE
% \begin{note}
%   To improve: put a bit into context? hard forks of Bitcoin and Ethereum?
% \end{note}

% Delegates can propose changes to part of the codebase
% and put this proposal to vote to other token holders. If the changes
% are adopted, nodes will run the for amending part of the code. The
% part of the code that is amendable by vote is referred to as the
% \emph{economic ruleset} or the \emph{economic
%   protocol}~\cite{goodman2014positionpaper,goodman2014whitepaper}. It
% consists of every rule that defines The voting procedures at the moments
% lasts about three months.
% % TODO shell / protocol division
% % emergency updates through hard forks
% % evthg that could be controversial is in the economic ruleset
% % todo add reference of whitepaper to Nomic

\subsubsection*{Focus on formal verification}
Our long-term ambition is to have certified code in the whole Tezos
codebase\footnote{Note that since code changes must be approved by the
  Tezos community, we can only propose a certified implementation of
  the economic ruleset.} as well as certified smart contracts.
%OCaml
The choice of OCaml as an implementation language is an interesting first step: OCaml gives Tezos good
static guarantees since it benefits from OCaml's strong type system
and memory management
features. % null pointer exceptions, buffer overflows
Furthermore, formally verified OCaml code can be produced by a variety
of tools such as F*~\cite{fstar}, Coq~\cite{coq}, Isabelle/HOL~\cite{isabelle}, Why3~\cite{why3}, and FoCaLiZe~\cite{focalize}. %
% This article is about a formalisation of the
%Cryptographic primitives
Another specificity of Tezos is the use of formally verified
cryptographic primitives. Indeed the codebase uses the HACL*
library~\cite{hacl}, which is certified C code extracted from an
implementation of Low*, a fragment of F*. %
This article presents Mi-Cho-Coq, a framework for formal verification
of Tezos smart contracts, written in the Michelson programming
language. It is organised as follows: Section~\ref{sec:michelson}
gives an overview of the Michelson smart contract language, the
Mi-Cho-Coq framework is then presented in
Section~\ref{sec:mi-cho-coq}, a case study on a Multisig smart
contract is then conducted in Section~\ref{sec:use-case-multisig},
Section~\ref{sec:related-work} presents some related workd and finally
Section~\ref{sec:limits-future-work} concludes the article by listing
directions for future work.

The Mi-Cho-Coq framework, including the Multisig contract
described in Section~\ref{sec:use-case-multisig}, is available at
\url{https://gitlab.com/nomadic-labs/mi-cho-coq/tree/FMBC_2019}.

% \subsubsection*{Smart contracts}
% The programming language for Tezos smart contracts is Michelson.

% Its smart contract language is called Michelson.
% % TODO develop account based and smart contract platform
% % and has an account based model, unlike Bitcoin~\cite{nakamoto2008bitcoin} which uses
% % Unspent Transactions Outputs (UTXO)~\cite{PrincetonBitcoinBook}. There are similar differences
% % though in the design. First the consensus algorithm relies on
% % Proof-of-Stake.

% Last but not least, Tezos has been designed with a strong focus on
% formal verification. OCaml's strong type system and memory management
% avoid many runtime errors. Furthermore, Tezos benefits from the
% cryptographic primitives of the HACL* library, implemented in the Low*
% fragment of F* and extracted to C.

%%% Local Variables:
%%% mode: latex
%%% TeX-master: "paper"
%%% End:

\section{Overview of Michelson}
\label{sec:michelson}
Smart contracts are Tezos accounts of a particular kind.  They have a
private access to a memory space on the chain called the
\emph{storage} of the smart contract, each transaction to a smart
contract account contains some data, the \emph{parameter} of the
transaction, and a \emph{script} is run at each transaction to decide
if the transaction is valid, update the smart contract storage, and
possibly emit new operations on the Tezos blockchain.

Michelson is the language in which the smart contract scripts are
written. The Michelson language has been designed before the launch of
the Tezos blockchain. The most important parts of the implementation
of Michelson, the typechecker and the interpreter, belong to the
economic ruleset of Tezos so the language can evolve through the Tezos
amendment voting process.

\subsection{Design rationale}
\label{sec:design-rational}

% constraints
Smart contracts operate in a very constrained context: they need to be
expressive, evaluated efficiently, and their resource consumption should be
accurately measured in order to stop the execution of programs that
would be too greedy, as their execution time impacts the block
construction and propagation.
Smart contracts are non-updatable programs that can handle valuable
assets, there is thus a need for strong guarantees on the correctness
of these programs.

%consequences
The need for efficiency and more importantly for accurate account of
resource consumption leans toward a low-level interpreted language, while the
need for contract correctness leans toward a high level, easily
auditable, easily formalisable language, with strong static
guarantees.

% design
To satisfy these constraints, Michelson was made a Turing-complete,
low level, stack based interpreted language (\textit{\`a la} Forth), enabling the resource
measurement, but with some high level features \textit{\`a la} ML:
polymorphic products, options, sums, lists, sets and maps data-structures with
collection iterators, cryptographic primitives and anonymous functions.
Contracts are pure functions that take a stack as input and return a stack as output. This side-effect free design is an asset
for the conception of verification tools.

The language is statically typed to ensure the well-formedness
of the stack at any point of the program. This means that if a
program is well typed, and if it is being given a well-typed stack that matches its
 input expectation, then at any point of the program execution, the
given instruction can be evaluated on the current stack.

Moreover, to ease the formalisation of Michelson, ambiguous or hidden
behaviours have been avoided. In particular, unbounded integers are
used to avoid arithmetic overflows and division returns an option
(which is !None! if and only if the divisor is 0) so that the
Michelson programmer has to specify the behaviour of the program in
case of division by 0; she can however still \emph{explicitly} reject
the transaction using the !FAILWITH! Michelson instruction.

\subsection{Quick tour of the language}
\label{sec:quick-tour-language}

The full language syntax, type system, and
semantics are documented in~\cite{michelsonwhitedoc}, we give
here a quick and partial overview of the language.

\subsubsection{Contracts' shape}
\label{sec:contracts-shape}

A Michelson smart contract script is written in three parts:
the parameter type, the storage type, and the code of the contract.
A contract's code consists in one block of code that can only be called with
one parameter, but multiple entry points can be encoded by branching on a
nesting of sum types and multiple parameters can be paired into one.

When the contract is deployed (or \emph{originated} in Tezos lingo) on the chain, it is bundled
with a data storage which can then only be changed by a contract successful
execution.
The parameter and the storage associated to the contract are paired
and passed to the contract's code at each execution, it has to return
a list of operations and the updated storage.

Seen from the outside, the type of the contract is the type of its
parameter, as it is the only way to interact with it.

\subsubsection{Michelson Instructions}
\label{sec:instruction-type}

As usual in stack-based languages, Michelson instructions take their parameters on the stack.
All Michelson instructions are typed as a function going from the expected
state of the stack, before the instruction evaluation, to the resulting
stack. For example, the !AMOUNT! instruction used to obtain the amount in $\mu tez$
of the current transaction has type !'S -> mutez:'S! meaning that
for any stack type !'S!, it produces a stack of type !mutez:'S!.
Some instructions, like comparison or arithmetic operations, exhibit
non-ambiguous ad-hoc polymorphism: depending on the input arguments' type,
a specific implementation of the instruction is selected, and the
return type is fixed.
For example !SIZE! has the following types:
\begin{tabular}[t]{lcl}
\begin{lstlisting}
bytes:'S -> nat:'S
string:'S -> nat:'S
\end{lstlisting}
&\hspace{3em} &
\begin{lstlisting}
set 'elt:'S -> nat:'S
map 'key 'val:'S -> nat:'S
list 'elt:'S -> nat:'S
\end{lstlisting}
\end{tabular}

While computing the size of a string or an array of bytes is similarly implemented, under the hood,
the computation of map size has nothing to do with the computation of
string size.

%Obviously, a sequence !instr1;instr2! where !instr1! has type !'A -> 'B!
%and !instr2! has type !'B -> 'C! will have type !'A -> 'C!.

Finally, the contract's code is required to take a stack
with a pair \emph{parameter}-\emph{storage}
and returns a stack with a pair \emph{operation list}-\emph{storage}:\\
!(parameter_ty*storage_ty):[] -> (operation list*storage_ty):[]!.

The operations listed at the end of the execution can change the
delegate of the contract, originate new contracts, or transfer tokens
to other addresses.  They will be executed right after the execution
of the contract.  The transfers can have parameters and trigger the
execution of other smart contracts: this is the only way to perform
\emph{inter-contract} calls.

\subsubsection{A short example - the Vote contract.}
\label{sec:short-example} %% As it's name implies, it allows users to vote for a candidate among a list set at origination. The contracts retains the number of votes towards each candidate. Any user can vote any number of time, but must send 5 tez (or $5000000\mu tez$) with each vote, otherwise the transaction is refused. Moreover, the contracts does not generate any transaction.

We want to allow users of the blockchain to vote for their favorite
formal verification tool. In order to do that, we create a
smart contract tasked with collecting the votes. We want any user to
be able to vote, and to vote as many times as they want, provided they
pay a small price (say 5 $tez$).
We originate the contract with the names of a selection of
popular tools: Agda, Coq, Isabelle and K\_framework, which are placed
in the long-term storage of the contract, in an associative map
between the tool's name and the number of registered votes (of course,
each tool starts with 0 votes).

In the figure \ref{fig:vote}, we present a voting contract, annotated
with the state of the stack after each line of code.  When actually
writing a Michelson contract, development tools (including an Emacs
Michelson mode) can interactively, for any point of the code, give the
type of the stack provided by the Michelson typecheck of a Tezos node.

\newbox\voting
\begin{lrbox}{\voting}
\begin{lstlisting}[numbers=left]
storage (map string int); # candidates
parameter string; # chosen
code {
  # (chosen, candidates):[]
  AMOUNT;  # amount:(chosen, candidates):[]
  PUSH mutez 5000000; COMPARE; GT;
  # (5 tez > amount):(chosen, candidates):[]
  IF { FAIL } {}; # (chosen, candidates):[]
  DUP; DIP { CDR; DUP };
  # (chosen, candidates):candidates:candidates:[]
  CAR; DUP; # chosen:chosen:candidates:candidates:[]
  DIP { # chosen:candidates:candidates:[]
        GET; ASSERT_SOME;
        # candidates[chosen]:candidates:[]
        PUSH int 1; ADD; SOME
        # (Some (candidates[chosen]+1)):candidates:[]
      }; # chosen:(Some (candidates[chosen]+1)):candidates:[]
  UPDATE; # candidates':[]
  NIL operation; PAIR # (nil, candidates'):[]
}
\end{lstlisting}
\end{lrbox}
\newbox\votingstorage
\begin{lrbox}{\votingstorage}
\begin{lstlisting}
{Elt "Agda" 0 ; Elt "Coq" 0 ; Elt "Isabelle" 0 ; Elt "K" 0}
\end{lstlisting}
\end{lrbox}
\begin{figure}[h!]
\centering
\captionsetup[subfigure]{position=b}
  \subcaptionbox
    {\label{fig:vote}}
    {\usebox\voting}
    \hfill
\centering
  \subcaptionbox
    {\label{fig:vote-storage}}
    {\usebox\votingstorage}
  % \begin{subfigure}{0.3\textwidth}
  %   \usebox\votingstorage
  %   \caption{\label{fig:vote-storage}}
  % \end{subfigure}
    \caption{A simple voting contract \subref{fig:vote} and an example of initial storage \subref{fig:vote-storage}}
\end{figure}

Let's take a look at our voting program: First, the description of the
storage and parameter types is given on lines \texttt{1-2}.  Then the
code of the contract is given.
On line \texttt{5}, !AMOUNT! pushes on the stack the amount
of (in $\mu tez$) sent to the contract address by the user. The threshold amount
(5 $tez$) is also pushed on the stack on line \texttt{6} and compared to
the amount sent:
!COMPARE! pops the two top values of the stack, and pushes
either -1, 0 or 1 depending on the comparison between the
value. !GT! then pops this value and pushes !true!
if the value is 1. If the threshold is indeed greater than the
required amount, the first branch of the !IF! is executed
and !FAIL! is called, interrupting the contract execution
and cancelling the transaction.

If the value was !false!, the execution continues on line
\texttt{9}, where we prepare the stack for the next action:
!DUP! copies the top of the stack, we then manipulate the
tail of the stack while preserving it's head using !DIP!:
there, we take the right element of the !(chosen, candidates)! pair
with !CDR!, and we duplicate it again.  By closing the block
guarded by !DIP! we recover the former stack's top, and the following
line takes its left element with !CAR!, and duplicates it.

On line \texttt{12}, we use !DIP! to protect the top of the
stack again. !GET! then pops !chosen! and
!candidates! from  the stack, and pushes an
option containing the number of votes of the candidate, if it was
found in the map. If it was not found, !ASSERT_SOME! makes
the program fail. On line \texttt{15}, the number of votes is
incremented by !ADD!, and packed into an option type by
!SOME!.

We then leave the !DIP! block to regain access to
value at the top of the stack (!chosen!). On line
\texttt{18}, !UPDATE! pops the three values remaining
on top of the stack, and pushes the !candidates! map updated
with the incremented value for !chosen!. Finally, we push an
empty list of operations with !NIL operation!, and pair the
two elements on top of the stack to get the correct return type.

%%% Local Variables:
%%% mode: latex
%%% TeX-master: "paper"
%%% End:

% LocalWords:  formedness

\section{Mi-Cho-Coq : a Verification Framework in Coq for Michelson}
\label{sec:mi-cho-coq}

Mi-Cho-Coq consists of an implementation of a Michelson interpreter in Coq as well as a
weakest precondition calculus à la Dijkstra~\cite{Dijkstra:1975:WP}.

\subsubsection{Michelson syntax and typing in Coq}

Michelson's type system, syntax and semantics, as described in the
main documentation, are fully formalised in Mi-Cho-Coq.

The abstract syntax tree of a Michelson script is a term of an inductive type
which carries the script type :
\lstset{language=Coq}
\begin{lstlisting}
Inductive instruction : list type -> list type -> Set :=
| NOOP {A} : instruction A A
| FAILWITH {A B a} : instruction (a :: A) B
| SEQ {A B C} : instruction A B -> instruction B C -> instruction A C
| IF {A B} : instruction A B -> instruction A B -> instruction (bool :: A) B
| LOOP {A} : instruction A (bool :: A) -> instruction (bool :: A) A ...
\end{lstlisting}

A Michelson code is usually a sequence of instructions (\coqe{SEQ}), which is
one of the \coqe{instruction} constructors.
It has type \coqe{instruction stA stB} where
\coqe{stA} and \coqe{stB} are respectively the type of the input stack
and of the output stack.

The stack type is a list of Michelson type constructions, defined in the
\coqe{type} inductive:
\begin{lstlisting}
Inductive comparable_type : Set :=
| nat | int | string | bytes | bool | mutez | address | key_hash | timestamp.

Inductive type : Set :=
| Comparable_type (a : comparable_type) | key | unit | signature | operation
| option (a : type) | list (a : type) | set (a : comparable_type)
| contract (a : type) | pair (a b : type) | or (a b : type) | lambda (a b : type)
| map (key : comparable_type) (val : type)
| big_map (key : comparable_type) (val : type).
\end{lstlisting}

A full contract, for a given storage type \coqe{storage} and parameter
type \coqe{params} is an instruction of type
\begin{lstlisting}
instruction ((pair params storage) :: nil) ((pair (list operation) storage) :: nil).
\end{lstlisting}

Thanks to the indexing of the ^instruction^ inductive by the input and
output stack types, only well-typed Michelson instructions are
representable in Mi-Cho-Coq. This is very similar to the
implementation of Michelson in the Tezos node which uses a similar
feature in OCaml: generalised algebraic datatypes.

To ease the transcription of Michelson contracts into Mi-Cho-Coq AST
we use notations so that contracts in Mi-Cho-Coq look very similar to
actual Michelson code. The main discrepancy between Michelson and
Mi-Cho-Coq syntax being that due to parsing limitations, the
Michelson semi-colon instruction terminator has to be replaced by a
double semi-colon instructions separator.

The ad-hoc polymorphism of Michelson instructions is handled by adding
an implicit argument to the corresponding instruction constructor in
Mi-Cho-Coq.
This argument is a structure that carries an element identifying the
actual implementation of the instruction to be used.
As the argument is \emph{implicit and maximally inserted}, Coq type
unifier tries to fill it with whatever value can fit with the known
types surrounding it, \emph{i.e.} the type of the input stack.
Possible values are declared through the Coq's canonical structures
mechanism, which is very similar to (Coq's or Haskell's) typeclasses.

\subsubsection{Michelson interpreter in Coq}

Michelson semantics is formalised in Coq as an evaluator \coqe{eval}
of type ^forall {A B : list type}, instruction A B -> nat -> stack A^ ^-> M (stack B)^
where ^M^ is the
error monad used to represent the explicit failure of the execution of
a contract. The argument of type ^nat^ is called the \emph{fuel} of the evaluator. It represents a bound on the depth
of the execution of the contract and should not be confused with
Michelson's cost model which is not yet formalised in
Mi-Cho-Coq.

Some domain specific operations which are hard to define in Coq are
axiomatised in the evaluator. These include cryptographic primitives,
data serialisation, and instructions to query the context of the call
to the smart contract (amount and sender of the transaction, current
date, balance and address of the smart contract).

\subsubsection{A framework for verifying smart contracts}

To ease the writing of correctness proofs in Mi-Cho-Coq, a weakest
precondition calculus is defined as a function \coqe{eval_precond} of
type ^forall {fuel A B}, instruction A B -> (stack B -> Prop) ->^ ^(stack A -> Prop)^
that is a Coq function taking as argument an instruction and a
predicate over the possible output stacks of the instruction (the
postcondition) and producing a predicate on the possible input stacks
of the instruction (the precondition).

This function is proved correct with respect to the evaluator:
\begin{lstlisting}
  Lemma eval_precond_correct {A B} (i : instruction A B) fuel st psi :
    eval_precond fuel i psi st <->
      match eval i fuel st with Failed _ _ => False | Return _ a => psi a end.
\end{lstlisting}

Note that the right-hand side formula is the result of the monad
transformer of \cite{fstar_monad_transformer} which here yields a
simple expression thanks to the absence of complex effects in
Michelson.

%%%%%%%%%%%%%%%%%%%%%%%%%%  partie écrite par Basile
\subsubsection{A short example - the Vote contract}
We give below, as an example, a formal specification of the voting
contract seen previously. We want the contract to take into account
every vote sent in a transaction with an amount superior to 5 $tez$.
Moreover, we want to only take into account the votes toward an actual
available choice (the contract should fail if the wrong name is sent
as a parameter). Finally, the contract should not emit any operation.

In the following specification, the \emph{precondition} is the
condition that must be verified for the contract to succeed. The
\emph{postcondition} fully describes the new state of the storage at the
end of the execution, as well as the potentially emitted
operations. \texttt{amount} refers to the quantity of $\mu tez$ sent
by the caller for the transaction.

\let\legacywedge\wedge
\def\wedge{\ensuremath{\legacywedge}}
{\small
\begin{longtable}{ll}
  \textbf{Precondition}: & \texttt{amount} $\geq$ 5000000 \wedge{} chosen $\in$ \texttt{Keys}(storage)\\
  \textbf{Postconditions}: & returned\_operations = [ ] \wedge\\
  & $\forall$ c, c $\in$ \texttt{Keys}(storage) $\iff$ c $\in$ \texttt{Keys}(new\_storage) \wedge\\
  & new\_storage[chosen] = storage[chosen] + 1 \wedge\\
  & $\forall$ c $\in$ \texttt{Keys}(storage), c $\neq$ chosen $\Rightarrow$ new\_storage[c] = storage[c]
\end{longtable}}

Despite looking simple, proving the correctness of the
vote contract still needs a fair number of properties about the map
data structure. In particular we need some lemmas about the relations
between the \texttt{mem}, \texttt{get} and \texttt{update} functions,
which we added to the Mi-Cho-Coq library to prove this contract.

Once these lemmas are available, the contract can easily be proved by studying the three different
situations that can arise during the execution : the contract can fail
(either because the sender has not sent enough tez or because they
have not selected one of the possible candidates), or the execution
can go smoothly. % The way we framed the precondition and postcondition of the contract reflects these three scenarios.

%%% Local Variables:
%%% mode: latex
%%% TeX-master: "paper"
%%% End:

\lstset{language=michelson}

\section{A case study : the Multisig Contract}
\label{sec:use-case-multisig}

The \textit{Multisig} contract is a typical example of access-control
smart contract. A Multisig contract is used to share the ownership of
an account between several owners. The owners are represented by their
cryptographic public keys in the contract storage and a pre-defined
\textit{threshold} (a natural number between 1 and the number of
owners) of them must agree for any action to be performed by the
Multisig contract.

Agreement of an owner is obtained by requiring a cryptographic
signature of the action to be performed.  To ensure that this
signature cannot be replayed by an attacker to authenticate in another
call to a Multisig contract (the same contract or another one
implementing the same authentication protocol), a nonce is appended to
the operation before signing. This nonce consists of the address of
the contract on the blockchain and a counter incremented at each call.

\subsubsection{Michelson Implementation}

% TODO: dans la partie Michelson, j’ai besoin que soient presentees
% les lambdas et les operations

To be as generic as possible, the possible actions of our Multisig
contract are:
\begin{itemize}
\item produce a list of operations to be run atomically
\item change the threshold and the list of owner public keys
\end{itemize}

% TODO: dans la partie Michelson, j’ai besoin de la notion de point
% d’entree (dire que c’est une application courante du type or).

The contract features two entrypoints named \texttt{default} and
\texttt{main}.  The \texttt{default} entrypoint takes no parameter (it
has type \texttt{unit}) and lets unauthenticated users send funds to
the Multisig contract. The \texttt{main} entrypoint takes as
parameters an action, a list of optional signatures, and a counter
value. It checks the validity and the number of signatures and, in
case of successful authentication, it executes the required action and
increment the counter.

The Michelson script of the Multisig contract is available at
\cite{multisigArthur}. The code of the \texttt{default} entrypoint
is trivial. The code for the \texttt{main} entrypoint can be divided
in three parts: the header, the loop, and the tail.

The header packs together the required action and the nonce and checks
that the counter given as parameter matches the one stored in the
contract.

The loop iterates over the stored public keys and the optional
signatures given in parameter. It counts and checks the validity of
all the signatures.

Finally the contract tail checks that the number of provided
signatures is at least as large as the threshold, it increments the
stored counter, and it runs the required action (it either evaluates
the anonymous function passed in the contract parameter and emits the
resulting operations or modifies the contract storage to update the
list of owner public keys and the threshold).

\subsubsection{Specification and Correctness Proof}

Mi-Cho-Coq is a functional verification framework. It is well suited
to specify the relation between the input and output stacks of a
contract such as Multisig but it is currently not expressive enough to
state properties about the lifetime of a smart contract nor the
interaction between smart contracts. For this reason, we have not
proved that the Multisig contract is resistant to replay
attacks. However, we fully characterise the behaviour of each call to
the Multisig contract using the following specification of the
Multisig contract, where !env! is the evaluation environment containing
among other data the address of the contract (!self env!) and the amount
of the transaction (!amount env!).

\lstset{language=Coq}
\begin{lstlisting}
Definition multisig_spec (parameter : data parameter_ty) (stored_counter : N)
           (threshold : N) (keys : Datatypes.list (data key))
           (new_stored_counter : N) (new_threshold : N)
           (new_keys : Datatypes.list (data key))
           (returned_operations : Datatypes.list (data operation))
           (fuel : Datatypes.nat) :=
  let storage : data storage_ty := (stored_counter, (threshold, keys)) in
  match parameter with
  | inl tt =>
    new_stored_counter = stored_counter /\ new_threshold = threshold /\
    new_keys = keys /\ returned_operations = nil
  | inr ((counter, action), sigs) =>
    amount env = (0 ~Mutez) /\ counter = stored_counter /\
    length sigs = length keys /\
    check_all_signatures sigs keys (fun k sig =>
         check_signature env k sig
           (pack env pack_ty (address_ env parameter_ty (self env),
                              (counter, action)))) /\
    (count_signatures sigs >= threshold)%N /\
    new_stored_counter = (1 + stored_counter)%N /\
    match action with
    | inl lam =>
      match (eval lam fuel (tt, tt)) with
      | Return _ (operations, tt) =>
        new_threshold = threshold /\ new_keys = keys /\
        returned_operations = operations
      | _ => False
      end
    | inr (nt, nks) =>
      new_threshold = nt /\ new_keys = nks /\ returned_operations = nil
    end end.
\end{lstlisting}

% Fixpoint check_all_signatures (sigs : Datatypes.list (Datatypes.option (data signature)))
%          (keys : Datatypes.list (data key))
%          (check_sig : data key -> data signature -> data bool) {struct keys} :=
%   match sigs, keys with
%   | nil, nil => true
%   | nil, cons _ _ => false
%   | cons _ _, nil => false
%   | cons (Some sig) sigs, cons k keys =>
%     andb (check_sig k sig) (check_all_signatures sigs keys check_sig)
%   | cons None sigs, cons _ keys =>
%     check_all_signatures sigs keys check_sig
%   end.

% Fixpoint count_signatures (sigs : Datatypes.list (Datatypes.option (data signature))) :=
%   match sigs with
%   | nil => 0%N
%   | cons None sigs => count_signatures sigs
%   | cons (Some _) sigs => (count_signatures sigs + 1)%N
%   end.

% Definition action_ty := or (lambda unit (list operation)) (pair nat (list key)).
% Definition pack_ty := pair address (pair nat action_ty).

% Verification fonctionelle complete mais pas de preuve de securite anti-rejeu

Using the Mi-Cho-Coq framework, we have proved the following theorem:

\begin{lstlisting}
Lemma multisig_correct (params : data parameter_ty)
      (stored_counter new_stored_counter threshold new_threshold : N)
      (keys new_keys : list (data key))
      (returned_operations : list (data operation)) (fuel : nat) :
  let storage : data storage_ty := (stored_counter, (threshold, keys)) in
  let new_storage : data storage_ty :=
    (new_stored_counter, (new_threshold, new_keys)) in
  17 * length keys + 14 $\leq$ fuel ->
  eval multisig (23 + fuel) ((params, storage), tt)
    = Return _ ((returned_operations, new_storage), tt) <->
  multisig_spec params stored_counter threshold keys
    new_stored_counter new_threshold new_keys returned_operations fuel.
  \end{lstlisting}

The proof relies heavily on the correctness of the precondition
calculus. The only non-trivial part of the proof is the signature
checking loop. Indeed, for efficiency reasons, the Multisig contract
checks the equality of length between the optional signature list and
the public key list only after checking the validity of the signature;
an optional signature and a public key are consumed at each loop
iteration and the list of remaining optional signatures after the loop
exit is checked for emptiness afterward. For this reason, the
specification of the loop has to allow for remaining unchecked
signatures.

% Some standard inductive properties of lists are used.

%%% Local Variables:
%%% mode: latex
%%% TeX-master: "paper"
%%% End:

\section{Related Work}
\label{sec:related-work}
Formal verification of smart contracts is a recent but active
field. The K framework has been used to formalise~\cite{kevm:2018} the
semantics of both low-level and high-level smart contract languages
for the Ethereum and Cardano blockchains.  These formalizations have
been used to verify common smart contracts such as Casper,
Uniswap, and various implentation of the ERC20 and ERC777 standards.

Ethereum smart contracts, written in the Solidity high-level language,
can also be certified using a translation to the F* dependently-typed
language~\cite{FVSC:2016}.

The Zen Protocol~\cite{zenprotocol_whitepaper} directly uses F* as its
smart contract language so that smart contracts of the Zen Protocol
can be proved directly in F*. Moreover, runtime tracking of resources
can be avoided since computation and storage costs are encoded in the
dependent types.

The Scilla~\cite{scilla2018} language of the Zilliqa blockchain has
been formalised in Coq. This language is higher-level (but also less
featureful) than Michelson. Its formalisation includes
inter-contract interaction and contract lifespan properties. This
has been used to show safety properties of a crowdfounding smart
contract.

% \item Archetype: high-level, state automata, use why3
% \item Cardano: maybe plutus or marlowe?
% plutus is haskell like
% plutus Core
% rebecca valentine, formal specification of the typed plutus core language
% \url{https://iohk.io/research/papers/#formal-specification-of-the-typed-plutus-core-language}
% \url{https://iohk.io/research/papers/#an-ontology-for-smart-contracts}

%%% Local Variables:
%%% mode: latex
%%% TeX-master: "paper"
%%% End:

\section{Limits and Future Work}
\label{sec:limits-future-work}
As we have seen, the Mi-Cho-Coq verification framework can be used to
certify the functional correctness of non-trivial smart contracts of
the Tezos blockchain such as the Multisig contract.  We are currently working
on several improvements to extend the expressivity of the framework;
Michelson's cost model and the semantics of inter-contract interactions are
being formalised.

In order to prove security properties, such as the absence of
signature replay in the case of the Multisig contract, an adversarial
model has to be defined. This task should be feasible in Coq but our
current plan is to use specialised tools such as
Easycrypt~\cite{easycrypt} and ProVerif~\cite{proverif}.

No code is currently shared between Mi-Cho-Coq and the Michelson
evaluator written in OCaml that is executed by the Tezos nodes. We
would like to raise the level of confidence in the fact that both
evaluators implement the same operational semantics. We could achieve
this either by proposing to the Tezos stakeholders to amend the
ecomomic protocol to replace the Michelson evaluator by a version
extracted from Mi-Cho-Coq or by translating to Coq the OCaml code of
the Michelson evaluator using a tool such as CoqOfOCaml~\cite{claret:phd} or CFML~\cite{cfml} and
then prove the resulting Coq function equivalent to the Mi-Cho-Coq
evaluator.

Last but not least, to ease the development of certified compilers
from high-level languages to Michelson, we are working on the design
of an intermediate compilation language called Albert that
abstracts away the Michelson stack.

%%% Local Variables:
%%% mode: latex
%%% TeX-master: "paper"
%%% End:

% Albert, reasoning on a lifetime of a contract


\bibliographystyle{splncs04}
\bibliography{../short_references}

\begin{thebibliography}{10}
\providecommand{\url}[1]{\texttt{#1}}
\providecommand{\urlprefix}{URL }
\providecommand{\doi}[1]{https://doi.org/#1}

\bibitem{michelsonwhitedoc}
{\relax Michelson: the language of Smart Contracts in Tezos}.
  \url{http://tezos.gitlab.io/mainnet/whitedoc/michelson.html}

\bibitem{tezosLiquidPos}
Proof-of-stake in \relax{Tezos}.
  \url{https://tezos.gitlab.io//mainnet/whitedoc/proof_of_stake.html}

\bibitem{tezosGitLab}
Tezos code repository. \url{https://gitlab.com/tezos/tezos}

\bibitem{tezosVoting}
Voting in \relax{Tezos}.
  \url{https://tezos.gitlab.io//mainnet/whitedoc/voting.html}

\bibitem{zenprotocol_whitepaper}
An introduction to the zen protocol.
  \url{https://www.zenprotocol.com/files/zen_protocol_white_paper.pdf} (2017)

\bibitem{fstar_monad_transformer}
Ahman, D., Hritcu, C., Mart{\'{\i}}nez, G., Plotkin, G.D., Protzenko, J.,
  Rastogi, A., Swamy, N.: Dijkstra monads for free. CoRR
  \textbf{abs/1608.06499} (2016), \url{http://arxiv.org/abs/1608.06499}

\bibitem{easycrypt}
Barthe, G., Dupressoir, F., Gr{\'{e}}goire, B., Kunz, C., Schmidt, B., Strub,
  P.: Easycrypt: {A} tutorial. In: Foundations of Security Analysis and Design
  {VII} - {FOSAD} 2012/2013 Tutorial Lectures. Lecture Notes in Computer
  Science, vol.~8604, pp. 146--166. Springer (2013).
  \doi{10.1007/978-3-319-10082-1\_6}

\bibitem{FVSC:2016}
Bhargavan, K., Delignat-Lavaud, A., Fournet, C., Gollamudi, A., Gonthier, G.,
  Kobeissi, N., Kulatova, N., Rastogi, A., Sibut-Pinote, T., Swamy, N.,
  Zanella-B{\'e}guelin, S.: Formal verification of smart contracts: Short
  paper. pp. 91--96. PLAS '16, ACM, New York, NY, USA (2016).
  \doi{10.1145/2993600.2993611}

\bibitem{proverif}
Blanchet, B.: Modeling and verifying security protocols with the applied pi
  calculus and proverif. Foundations and Trends in Privacy and Security
  \textbf{1}(1-2),  1--135 (2016). \doi{10.1561/3300000004}

\bibitem{multisigArthur}
Breitman, A.: {\relax Multisig contract in Michelson}.
  \url{https://github.com/murbard/smart-contracts/blob/master/multisig/michelson/generic_multisig.tz}

\bibitem{cfml}
Chargu{\'e}raud, A.: Characteristic formulae for the verification of imperative
  programs. pp. 418--430. ICFP '11, ACM, New York, NY, USA (2011).
  \doi{10.1145/2034773.2034828}

\bibitem{claret:phd}
Claret, G.: {Program in Coq}. Theses, {Universit{\'e} Paris Diderot - Paris 7}
  (Sep 2018), \url{https://hal.inria.fr/tel-01890983}

\bibitem{Dijkstra:1975:WP}
Dijkstra, E.W.: Guarded commands, nondeterminacy and formal derivation of
  programs. Commun. ACM  \textbf{18}(8),  453--457 (Aug 1975).
  \doi{10.1145/360933.360975}

\bibitem{why3}
Filli{\^a}tre, J.C., Paskevich, A.: {Why3 -- Where Programs Meet Provers}. In:
  {ESOP'13 22nd European Symposium on Programming}. LNCS, vol.~7792.
  {Springer}, Rome, Italy (Mar 2013), \url{https://hal.inria.fr/hal-00789533}

\bibitem{goodman2014positionpaper}
Goodman, {\relax L.M}.: Tezos: A self-amending crypto-ledger. position paper.
  \url{https://tinyurl.com/tezospp} (2014)

\bibitem{goodman2014whitepaper}
Goodman, {\relax L.M}.: Tezos: A self-amending crypto-ledger. white paper.
  \url{https://tinyurl.com/tezoswp} (2014)

\bibitem{kevm:2018}
Hildenbrandt, E., Saxena, M., Zhu, X., Rodrigues, N., Daian, P., Guth, D.,
  Moore, B., Zhang, Y., Park, D., \c{S}tef\u{a}nescu, A., Ro\c{s}u, G.: Kevm: A
  complete semantics of the ethereum virtual machine. In: 2018 IEEE 31st
  Computer Security Foundations Symposium. pp. 204--217. IEEE (2018)

\bibitem{leroy:ocaml4.08}
Leroy, X., Doligez, D., Frisch, A., Garrigue, J., R{\'e}my, D., Vouillon, J.:
  {The OCaml system release 4.08: Documentation and user's manual}. User
  manual, {Inria} (Jun 2019), \url{http://caml.inria.fr/pub/docs/manual-ocaml/}

\bibitem{isabelle}
Nipkow, T., Wenzel, M., Paulson, L.C.: Isabelle/HOL: A Proof Assistant for
  Higher-order Logic. Springer-Verlag, Berlin, Heidelberg (2002)

\bibitem{focalize}
Pessaux, F.: {FoCaLiZe: Inside an F-IDE}. In: {Workshop F-IDE 2014}.
  Proceedings F-IDE 2014, Grenoble, France (May 2014).
  \doi{10.4204/EPTCS.149.7},
  \url{https://hal.archives-ouvertes.fr/hal-01203501}

\bibitem{scilla2018}
Sergey, I., Kumar, A., Hobor, A.: Scilla: a smart contract intermediate-level
  language. CoRR  \textbf{abs/1801.00687} (2018),
  \url{http://arxiv.org/abs/1801.00687}

\bibitem{fstar}
Swamy, N., Hritcu, C., Keller, C., Rastogi, A., Delignat-Lavaud, A., Forest,
  S., Bhargavan, K., Fournet, C., Strub, P.Y., Kohlweiss, M., Zinzindohou\'e,
  J.K., {Zanella-B\'eguelin}, S.: Dependent types and multi-monadic effects in
  {F*}. In: POPL. pp. 256--270. ACM (Jan 2016),
  \url{https://www.fstar-lang.org/papers/mumon/}

\bibitem{coq}
{The {Coq} development team}: {The {Coq} Reference Manual, version 8.9} (Nov
  2018), \url{http://coq.inria.fr/doc}

\bibitem{hacl}
Zinzindohou\'e, J.K., Bhargavan, K., Protzenko, J., Beurdouche, B.: Hacl*: A
  verified modern cryptographic library. Cryptology ePrint Archive, Report
  2017/536

\end{thebibliography}

% \appendix

% \section{Michelson Script of the Multisig Contract}
% \label{sec:multisig_appendix}
% \lstinputlisting{multisig.tz}

% \section{(REMOVE) Draft - ideas}
% \input{draft}

\end{document}